\documentclass[se, manuscript]{copernicus}

\usepackage[version=4]{mhchem}
\usepackage{siunitx}
\usepackage{xcolor}
\newcommand{\cel}[1]{\SI{#1}{\celsius}}
\newcommand{\GPa}[1]{\SI{#1}{\giga\pascal}}

\AtBeginDocument{\nolinenumbers}

\begin{document}

\title{Modeling of reactive porosity waves as a fluid release process in a dehydrating slab}

\Author[1][liudmila.khakimova@unil.ch]{Liudmila} {Khakimova}
\Author[2][]{Konstantin}{Huber}
\Author[1][]{Yury Y.}{Podladchikov}{}
\Author[2][]{Johannes C.}{Vrijmoed}{}
\Author[2][]{Boris}{Antonenko}{}
\Author[2][]{Timm}{John}{}

\affil[1]{Institute of Earth Sciences, University of Lausanne, Switzerland}
\affil[2]{Institute of Geological Sciences, Free University Berlin, Germany}




\runningtitle{Reactive porosity waves in a dehydrating slab}

\runningauthor{Khakimova et al.}

\received{}
\pubdiscuss{} 
\revised{}
\accepted{}
\published{}


\firstpage{1}

\maketitle

\begin{abstract}
Dehydration of subducting serpentinites releases aqueous fluids that must be efficiently extracted from the slab. We present a fully coupled two-dimensional hydro--mechanical--chemical model of reactive porosity waves in deforming and dehydrating serpentinite. Thermodynamic closure is provided by Gibbs free-energy minimization for the MgO--FeO--SiO$_2$--H$_2$O system, including dissolved SiO$_2$ in the aqueous phase.
Natural multiscale compositional maps of serpentinites
from the Mirdita ophiolite (Albania) are used to prescribe initial SiO$_2$ and FeO heterogeneities.
The simulations show that these intrinsic chemical variations trigger spontaneous localization of porosity and fluid flow during dehydration, without prescribing channel geometry.
Channelization results from the coupling of decompaction, fluid-pressure buildup, buoyancy-driven flow, and metasomatic reactions associated with infiltration of SiO$_2$-poor hydrous fluid and olivine formation. The resulting fluid-rich channels propagate upward and develop connected networks resembling natural vein systems. Our results demonstrate that reactive porosity waves in compositionally heterogeneous serpentinites provide a robust mechanism for focusing fluid flow and enhancing fluid extraction from dehydrating subducting slabs.
\end{abstract}

\section{Introduction}

Subduction zones are among Earth’s most dynamic systems, where seawater-altered oceanic lithosphere descends into the mantle. They link Earth’s exogenic and endogenic reservoirs and strongly influence the global chemical and energy budgets. As the subducting plate heats up, hydrous minerals progressively destabilize and release their bound water. Slab descent rates of several centimeters per year require efficient fluid extraction mechanisms to prevent complete  loss of water into the deep mantle \citep{van2011subduction, plumper2017}. Therefore, the efficiency of slab devolatilization is crucial for maintaining the overall stability of the Earth's volatile reservoirs over geological timescales. The inherently low permeabilities of subducting rocks strongly limit pervasive, Darcy-type flow, making such distributed fluid release inefficient \citep{connolly1998, plumper2017}. Instead, channelised flow is widely recognised as the dominant mechanism for high-flux fluid transport through the slab \citep{zack2007, john2012, plumper2017, taetz2018, tian2019devolatilization, piccoli2021}.

The descending slab densifies during progressive dehydration-related metamorphism, while the released fluids alter the physical and chemical properties of the rocks they pass through. Fluid–rock interaction can overprint mineral assemblages, facilitate deformation, alter seismic wave speeds, and create chemical alteration zones \citep{john2004,gao2007,rondenay2008seismic,spandler2011,chen2019,zertani2019modification, kaatz2022, yamato2022reaction}. 
Fluids released during slab dehydration can reach the mantle wedge and may be directly linked to mantle-wedge seismicity and the cause of melting \citep[e.g.][]{mcgary2014,halpaap2019earthquakes,white2019earth,yoshida2023}.
The chemical freight of these fluxing fluids is thought to create the distinct chemical signatures of the resulting arc magmas \citep[e.g.][]{elliott2003,john2012,spandler2013element,chen2019,plank2019subducting,li2020}.

Recent numerical work has shown that intrinsic chemical heterogeneities and reactive flow can strongly influence the onset of dehydration and promote the evolution of interconnected vein networks \citep{huber2022,huber2024}. Yet dehydration reactions operate on micrometer- to millimeter-scales, while larger-scale gravitational effects govern fluid segregation \citep{plumper2017} if the porous media interacts with the free fluid (two phase flow). Incorporating two-phase effects introduces buoyancy-driven deformation of the porous medium: compaction expels fluids and reduces porosity, whereas decompaction allows fluid accumulation and upward migration.

Fluid expulsion can occur through several mechanisms. In brittle regimes, fluid overpressure may exceed the rock's tensile strength and induce hydraulic fracturing, generating transient, highly permeable conduits that accommodate rapid drainage \citep{miller2003,bons2012review,yoshida2023}. In more ductile conditions, fluid transport may be governed by porosity waves, self-sustaining zones of localized high porosity driven by fluid overpressure in visco-elastoplastic rocks \citep{connolly1998,miller2003,rass2014,wilson2014,rass2018}. At the leading edge of such waves, fluid overpressure induces decompaction, facilitating upward propagation; at the trailing edge, pressure drops induce compaction, closing porosity. Rock rheology exerts a first-order control on wave morphology: decompaction‐weakening, where decompaction occurs more readily than compaction, promotes channeling instabilities and elongated porosity structures \citep{connolly2007,connolly2015, yarushina2015a, rass2019,yarushina2020}. These mechanisms are increasingly invoked to explain natural permeability structures such as fluid chimneys in sedimentary basins {}\citep{rass2018, yarushina2022}, fluid-escape pipes associated with \ce{CO2} storage \citep{johnson2024}, but also petrological observations of a kilometer-sized, high-pressure metasomatic system in an inferred paleo-subduction zone \citep{piccoli2021}.

Fluid flow in subduction zones is fundamentally reactive \citep{john2004,spandler2011,vanderstraaten2008,ague2014carbon,ague2022slab,manning2020subduction}. Chemical heterogeneities influence the composition of dehydration fluids, and because fluid and solid phases continuously attempt to achieve thermodynamic equilibrium, compositional variability directly feeds back on reaction progress and fluid chemistry. Reactive flow can change mineralogy, bulk composition, porosity, and permeability of the rock matrix \citep[e.g.][]{huber2022,bras2023eclogitisation, schmalholz2024hydration, schmalholz2026channelization}. Mineral reactions alter solid density, while dissolution–precipitation reactions can strongly modify fluid density \citep{john2012}. Capturing these coupled processes to finally derive a more complete understanding of the feedback among slab dehydration, intraslab seismicity and arc magmatism requires models that integrate thermodynamics, mass transport, and mechanics \citep[e.g.][]{bessat2022,khakimova2024,khakimova2025high,porkolab2025modeling}.

The model presented here constitutes an important step toward such integrated descriptions of rock deformation, fluid migration and element transport. It incorporates two key advances over previous approaches. First, mechanics and thermodynamic data are fully coupled using thermodynamic lookup tables that track fluid and rock chemistry, mineral reactions, density changes, and volumetric responses across the relevant P–T–X space. Second, the model explicitly includes realistic chemical heterogeneities by incorporating compositional maps derived from natural rock samples \citep{huber2024}. Although implementing such steep and spatially variable compositional gradients is numerically challenging, recent improvements in the numerical scheme allow simulations to run efficiently without requiring high-performance computing facilities \citep{khakimova2025reactive}.

Combining fully coupled hydro-mechanical-chemical (HMC) models with natural rock compositions opens new opportunities to investigate the dynamic evolution of porosity during rock dehydration. This integrated approach allows us to explore how chemical heterogeneity, reactive flow, mechanical, and gravitational processes interact to produce channelised fluid pathways. This is a key step toward understanding fluid release, seismicity, and magma generation in subduction zones.

\section{Model formulation}

\begin{table}[t]
\centering
\caption{Notations and shorthand definitions.}
\label{tab:notation}
\renewcommand{\arraystretch}{1.2}
\begin{tabular}{lll}
\hline
\textbf{Symbol} & \textbf{Meaning} & \textbf{Unit} \\
\hline
$P^s,\,P^f$ 
& Solid, and fluid pressures 
& Pa \\

$P^t$
& Total pressures, $P^f\varphi + P^s(1-\varphi) $
& Pa \\

$\tau^s,\,\tau^f$ 
& Solid and fluid deviatoric stress tensors 
& Pa \\

$\tau^t$ 
& Total deviatoric stress tensors, $\tau^f\varphi + \tau^s(1-\varphi)$ 
& Pa \\

$v^s,\,v^f$ 
& Solid and fluid particle velocities 
& m\,s$^{-1}$ \\

$q$ 
& Darcy flux, $\varphi(v^f - v^s)$ 
& m\,s$^{-1}$ \\

$K^s,\,K^f$ 
& Solid and fluid bulk moduli 
& Pa \\

$G$ 
& Shear modulus 
& Pa \\


$\eta_s$ 
& Shear viscosity of the solid matrix 
& Pa\,s \\

$\eta_\varphi$ 
& Bulk (compaction) viscosity of the porous rock  
& Pa\,s \\

$k / \eta_f$ 
& Permeability divided by fluid viscosity 
& m$^{2}$\,Pa$^{-1}$\,s$^{-1}$ \\

$R$ 
& Decompaction weakening factor 
& -- \\

$\varphi$ 
& Porosity 
& -- \\


$\rho^s,\,\rho^f$ 
& Solid and fluid mass densities 
& kg\,m$^{-3}$ \\

$\rho^t$ 
& Total density, $\rho^f\varphi + \rho^s(1-\varphi)$ 
& kg\,m$^{-3}$ \\

$C^{f,SiO_2}$ 
& Mass concentration of the $\mathrm{SiO_2}$ component in the fluid
& --  \\

$C^{s,k}$ 
& Mass concentration of the component $k=\{\mathrm{SiO_2},\, \mathrm{MgO},\, \mathrm{FeO}\}$ in the solid
& -- \\

$C^{s,im}$ 
& Mass concentration of the immobile (non-reacting) part of the solid matrix, $C^{s,MgO} + C^{s,FeO}$
& --  \\

$\rho C^{t,SiO_2}$ 
& Mass of the $\mathrm{SiO_2}$ component per unit volume in the system, $\rho^f\varphi C^{f,SiO_2} + \rho^s(1-\varphi) C^{s,SiO_2}$
& kg\,m$^{-3}$ \\

$\rho C^{t,k}$ 
& Mass of the immobile component $k=\{\mathrm{MgO},\, \mathrm{FeO}\}$ per unit volume in the system, $\rho^s(1-\varphi) C^{s,k}$
& kg\,m$^{-3}$ \\



$T$ 
& Temperature of the saturated rock
& K \\

$\delta_{ij}$ 
& Kronecker delta 
& -- \\

$\dfrac{d^s}{dt}$ 
& Material derivative with respect to the solid phase, 
$\displaystyle \frac{\partial}{\partial t} + v_i^s \nabla_i$ 
& -- \\

$\dfrac{d^f}{dt}$ 
& Material derivative with respect to the fluid phase, 
$\displaystyle \frac{\partial}{\partial t} + v_i^f \nabla_i$ 
& -- \\

$\dfrac{d^\nabla \tau_{ij}}{dt}$ 
& Jaumann stress rate for the solid matrix
& Pa\,s$^{-1}$ \\

$\omega_{ij}$ 
& Antisymmetric part of the solid velocity gradient, 
$\displaystyle \omega_{ij} = \tfrac{1}{2}(\nabla_j v_i^s - \nabla_i v_j^s)$ 
& s$^{-1}$ \\

\hline
\end{tabular}
\end{table}

We consider infiltration of a one-phase supercritical two-component fluid into a deformable and chemically reactive viscoelastic porous medium. The system contains four chemical components, $\{\mathrm{H_2O},\, \mathrm{SiO_2},\, \mathrm{MgO},\, \mathrm{FeO}\}$. Of these, $\mathrm{MgO}$ and $\mathrm{FeO}$ are classified as immobile components that remain structurally bound to the solid matrix over the relevant range of pressure, temperature and concentration of the system. The remaining components, $\mathrm{H_2O}$ and $\mathrm{SiO_2}$, are mobile and may partition between the solid and fluid phases, thereby participating in advective and diffusive transport and chemical mass exchange. 

The governing closed system of equations is presented below and derived from balance laws for the mass and momentum of each phase and each chemical component, formulated within the framework of classical thermodynamics. Under the assumption of local thermodynamic equilibrium in the weak formulation, following the approach of \citet{yarushina2015}, one obtains constitutive relations that ensure the thermodynamic consistency of the coupled hydro-mechanical-chemical (HMC) model. This methodology for deriving self-consistent continuum-scale equations originates from the classical principles outlined by \citet{landau1987fluid} and has been systematically extended in recent studies \citep{bessat2022, yarushina2023constitutive, khakimova2024, schmalholz2023, schmalholz2024hydration, schmalholz2026channelization}. A complete list of notation and shorthand definitions used in the model formulation is provided in Table~\ref{tab:notation}.

\subsection{Conservation laws and constitutive relations}
We begin with the conservation of total mass for the coupled fluid-solid system,
\begin{equation}
    \frac{\partial \rho^t}{\partial t}
    =
    - \nabla_i
    \left[
        (v_i^f - v_i^s)\,\rho^f \phi
        + v_i^s \rho^t
    \right],
    \label{eq:rpw_totmascon}
\end{equation}
where superscripts $s$ and $f$ denote the solid and fluid phases, respectively, and the subscript $i$ refers to spatial coordinates $i=(x,y,z)$, with the Einstein summation convention implied over repeated indices. The total mass density $\rho^t$ is defined as $\rho^t = \rho^f \phi + \rho^s (1-\phi)$, which represents the total mass of the two-phase system per unit volume.

Under the assumption that $\mathrm{MgO}$ and $\mathrm{FeO}$ are immobile and cannot partition into the fluid phase, their mass is conserved within the solid matrix. Consequently, the mass conservation equation for the immobile solid in unit volume can be written, without source term, as:
\begin{equation}
    \frac{\partial}{\partial t}\left[\rho^s (1-\varphi) C^{s,im} \right] = -\nabla_i \left[  v_i^s \rho^s (1-\varphi) C^{s,im} \right],
    \label{eq:rpw_immobmasscon}
\end{equation}
where the shortcut notation $C^{s,\mathrm{MgO}} + C^{s,\mathrm{FeO}} = C^{s,\mathrm{im}}$ denotes the mass concentration of the nonvolatile fraction of the solid matrix, which consists of the immobile components $\mathrm{MgO}$ and $\mathrm{FeO}$.

The mass conservation equations for the mobile $\mathrm{SiO_2}$ component, neglecting diffusive fluxes, can be written as
\begin{equation}
\frac{\partial}{\partial t}\!\left[ \rho\, C^{t,SiO_2} \right]
= - \nabla_i
\left[
    v^f \,\phi\,\rho^f\,C^{f,SiO_2}
    + v^s\ (1-\phi) \rho^s\,C^{s,SiO_2}
\right],
    \label{eq:total_components_conservation}
\end{equation}
where $C^{f,SiO_2}$ and $C^{s,SiO_2}$ are the mass concentrations of $\mathrm{SiO_2}$ in the fluid and solid phases, respectively, and $\rho\,C^{t,SiO_2}$ is the total mass of component $\mathrm{SiO_2}$ per unit volume of the system, which is defined in Table \ref{tab:notation}.


The fluid momentum conservation is given by Darcy's law in the form of 
\begin{equation}
    \phi \left( v_i^f - v_i^s \right) = - \frac{k}{\eta^f} \left( \nabla_i P^f + g_i \rho^f \right)
    \label{eq:rpw_Darcy}
\end{equation}
with $k$ being permeability and $\eta^f$ fluid viscosity. 
Permeability is linked to porosity by the Carman-Kozeny relationship 
\begin{equation}
    k = k^0 \left( \frac{\phi}{\phi^0} \right)^3
\end{equation}
with $k^0$ and $\phi^0$ being the background permeability and background porosity, respectively. 

The total momentum conservation is given by 
\begin{equation}
    \nabla_j \left( - P^t \delta_{ij} + \tau^t_{ij}\right) - g_i \rho^{t} = 0. 
    \label{eq:rpw_totmomcon}
\end{equation}

The constitutive relation linking the deviatoric stress tensor to the gradients of the solid velocity field, corresponding to a Maxwell viscoelastic rheology:
\begin{equation}
\frac{1}{2G}\,\frac{d^{\nabla} \tau_{ij}}{dt}
+ \frac{\tau_{ij}}{2\eta^s}
=
\frac{1}{2}
\left(
\nabla_j v_i^s + \nabla_i v_j^s
\right)
-
\frac{1}{3}
\left(
\nabla_k v_k^s
\right)\delta_{ij},
\label{eq:maxwell_deviatoric}
\end{equation}
where $G$ — shear modulus, $\eta^s$ — shear viscosity of the solid matrix, $d^{\nabla}/dt$ — Jaumann derivative, $\tau_{ij}$ — deviatoric stress tensor, and $v_i^s$ — solid velocity components.

The bulk viscosity, $\eta^{\varphi}$, is inversely proportional to porosity and is reduced by a factor $R$ in regions of fluid overpressure to account for decompaction weakening \citep{connolly1998, connolly2007, rass2019, yarushina2020}, here parametrized by a hyperbolic tangent function:

\begin{equation}
\eta_{\varphi} = \eta_c\,\frac{\phi_0}{\phi}
\left[
1 + \frac{1}{2}\left(\frac{1}{R}-1\right)
\left(1+\tanh\left[\frac{P^f - P^t}{\lambda}\right]\right)
\right],
\label{eq:eta_phi}
\end{equation}
where $\lambda$ is the transition-zone sharpness, and the reference bulk compaction viscosity, $\eta_c$, at reference porosity $\phi_0$ and $P^t \gg P^f$, is:
\begin{equation}
\eta_c = \frac{\eta_s}{C\,\phi_0},
\label{eq:eta_c}
\end{equation}
where $C$ is a rheological constant quantifying the ratio of shear viscosity over the reference bulk compaction viscosity.

Using the adopted rheological relations, the elastic compressibilities of the fluid and solid phases can be written as \citep{yarushina2015}:
\begin{equation}
\frac{K^f}{\rho^f}\,\frac{d^f \rho^f}{dt}
=
\frac{d^f P^f}{dt},
\label{eq:EOS_rhof}
\end{equation}

\begin{equation}
\frac{K^s}{\rho^s}\,\frac{d^s \rho^s}{dt}
=
\frac{d^s P^s}{dt},
\label{eq:EOS_rhos}
\end{equation}
where $K^f$ and $K^s$ are the bulk moduli of the fluid and solid phases, respectively, and $d^f/dt$ and $d^s/dt$ denote material derivatives taken with respect to the fluid and solid phases.

Using the relation between fluid, solid, and total pressure
\begin{equation}
P^t = \varphi\,P^f + (1-\varphi)\,P^s,
\label{eq:Pt}
\end{equation}
Equation~\eqref{eq:EOS_rhos} can be rewritten as
\begin{equation}
\frac{K^s}{\rho^s}\,\frac{d^s \rho^s}{dt}
=
\frac{1}{1-\varphi}
\left(
\frac{d^s P^t}{dt}
-
\varphi\,\frac{d^s P^f}{dt}
+
\frac{P^t-P^f}{1-\varphi}\,\frac{d^s\varphi}{dt}
\right).
\label{eq:EOS_rhos2}
\end{equation}

To close the system of governing equations, we employ an equation for porosity evolution derived from the viscous compaction of the solid matrix \citep{yarushina2015}. Under the assumption that volumetric deformation of the porous skeleton is controlled by viscous rheology, the porosity evolution is directly linked to the divergence of the solid velocity field and takes the following form:
\begin{equation}
    \frac{d^s \varphi}{dt} = \frac{P^f - P^t}{\eta^\varphi},
    \label{eq:rpw_viscous_rheologoy}
\end{equation}
where $\eta^\varphi$ is the bulk viscosity of the porous media. 

In the 2D configuration considered in this work, the governing system comprises 13 equations for 15 unknowns: $P^f$, $P^t$, $\rho\,C^{t,\mathrm{SiO_2}}$, $v_x^f$, $v_y^f$, $v_x^s$, $v_y^s$, $\tau_{xx}$, $\tau_{yy}$, $\tau_{xy}$, $\varphi$, $\rho^f$, $\rho^s$, $C^{s,\mathrm{im}}$, and $C^{f,\mathrm{SiO_2}}$.
In our implementation, we solve:\\
1) the total mass conservation equation \eqref{eq:rpw_totmascon} for the fluid pressure $P^f$;\\
2) the mass conservation equation for the immobile solid \eqref{eq:rpw_immobmasscon} for the total pressure $P^t$;\\
3) the total mass conservation equation for the mobile $\mathrm{SiO_2}$ component \eqref{eq:total_components_conservation} for $\rho\,C^{t,\mathrm{SiO_2}}$;\\
4--5) the fluid momentum balance \eqref{eq:rpw_Darcy} for the two components of fluid velocity, $v_x^f$ and $v_y^f$;\\
6--7) the total momentum balance \eqref{eq:rpw_totmomcon} for the two components of solid velocity, $v_x^s$ and $v_y^s$;\\
8--10) the Maxwell viscoelastic rheology \eqref{eq:maxwell_deviatoric} for the three independent components of the deviatoric stress tensor, $\tau_{xx}$, $\tau_{yy}$, and $\tau_{xy}$;\\
11) the porosity evolution equation with a viscous-compaction closure (\eqref{eq:rpw_viscous_rheologoy}), described by \citet{malvoisin2015} as a “classic form” and later discussed by \citet{yarushina2023constitutive}, to update the porosity $\varphi$;\\
12) the fluid compressibility relation \eqref{eq:EOS_rhof} for the fluid density $\rho^f$;\\
13) the solid compressibility relation \eqref{eq:EOS_rhos2} for the solid density $\rho^s$.

Despite this, the system is not yet closed because additional closure relations are still required to specify $C^{s,\mathrm{im}}$, and $C^{f,\mathrm{SiO_2}}$. For completeness, we also present the fluid and solid equations of state (last two equations listed above) in this section to maintain consistency with the model formulation of \citet{yarushina2015} derived for coupled hydro-mechanical system. However, because we extend the model to a coupled hydro–mechanical–chemical system, we do not use the previous constitutive laws; instead, the phase densities (as well as concentrations $C^{s,\mathrm{im}}$, and $C^{f,\mathrm{SiO_2}}$) are obtained directly from thermodynamic equilibrium calculations. These quantities provide the required thermodynamic closure and are implemented via the lookup tables described in the next section.

\subsection{Thermodynamic closure relationships}

Since the governing system remains underdetermined, with more unknowns than independent equations, we close the formulation by employing a lookup table strategy.

In this approach, the thermodynamic quantities that cannot be solved for
explicitly are precomputed via Gibbs free energy minimization and stored as
discrete entries in multidimensional arrays. Each table entry corresponds to the
outcome of a single Gibbs energy minimization, from which all equilibrium
quantities are subsequently extracted. These tables therefore serve as
constitutive relations, providing functional mappings for the unknown variables
in terms of those directly obtained from the conservation equations.

In contrast to analytical constitutive laws, the lookup-table approach provides discrete mappings obtained directly from thermodynamic equilibrium calculations. During numerical simulations, the required thermodynamic quantities are recovered by multidimensional interpolation of the tabulated data. In the present study, the lookup tables supply the fluid-phase silica concentration $C^{f,\mathrm{SiO_2}}$, the concentration of the immobile solid fraction $C^{s,\mathrm{im}}$, and the intrinsic densities of the solid and fluid phases $\rho^s$ and $\rho^f$:
\begin{align}
\label{eq:1rpw_LUT1}
C^{f,\mathrm{SiO_2}} &=
C^{f,\mathrm{SiO_2}}
\bigl(\Tilde{C}^{s,\mathrm{SiO_2}},\, \Tilde{C}^{s,\mathrm{MgO}},\, P^f, T\bigr), \\
\label{eq:1rpw_LUT2}
C^{s,\mathrm{im}} &=
C^{s,\mathrm{im}}
\bigl(\Tilde{C}^{s,\mathrm{SiO_2}},\, \Tilde{C}^{s,\mathrm{MgO}},\, P^f, T\bigr), \\
\label{eq:1rpw_LUT3}
\rho^s &=
\rho^s
\bigl(\Tilde{C}^{s,\mathrm{SiO_2}},\, \Tilde{C}^{s,\mathrm{MgO}},\, P^f, T\bigr), \\
\label{eq:1rpw_LUT4}
\rho^f &=
\rho^f
\bigl(\Tilde{C}^{s,\mathrm{SiO_2}},\, \Tilde{C}^{s,\mathrm{MgO}},\, P^f, T\bigr),
\end{align}
where equations~\eqref{eq:1rpw_LUT3} and \eqref{eq:1rpw_LUT4} are used in place of the constitutive density relations previously employed for hydro-mechanical systems (equations~\eqref{eq:EOS_rhof} and \eqref{eq:EOS_rhos2}). Here, all tabulated quantities are expressed as functions of the component mass concentrations in the solid matrix, $C^{s,\mathrm{SiO_2}}$ and $C^{s,\mathrm{MgO}}$,the fluid pressure $P^f$, and temperature $T$. Note, that within this work we assume the isothermal case and provide all calculation at constant temperature. 

\section{Numerical implementation and modeling setup}

The governing system constitutes a fully coupled hydro–mechanical–chemical problem, in which fluid flow, deformation, and reactive mass transport evolve simultaneously. Solving such a system requires coupling of a transport solver with thermodynamic equilibrium calculations. 

During the thermodynamic preprocessing stage, the functions given by equations
\eqref{eq:1rpw_LUT1}-\eqref{eq:1rpw_LUT4} and required by the transport model are computed and tabulated following \citet{connolly2005}. At each time step of the transport solver, these functions (equations \eqref{eq:1rpw_LUT1}-\eqref{eq:1rpw_LUT4}) are evaluated by interpolation from the precomputed multidimensional lookup tables. The lookup tables themselves are generated in a preprocessing step by performing Gibbs free-energy minimization (via linear programming) over the relevant ranges of state variables for every grid point in composition–pressure–temperature space. This procedure yields the equilibrium phase assemblages and their properties, which serve as the constitutive relations for the transport solver.

To consistently couple thermodynamic data computed in physical units with the HMC transport solver and to ensure numerical robustness, we introduce a set of dimensionally independent scales based on the domain length $L_x$, shear viscosity $\eta^s$ and the buoyancy force per unit volume $\Delta\rho g$, where $\Delta\rho = \rho^s-\rho^s$. These choices define the fundamental scales for length, pressure, and time, from which all remaining dimensional variables can be scaled. The governing equations are then nondimensionalized accordingly, and all physical variables in the code are renormalized using these reference scales. All thermodynamic quantities obtained from equilibrium calculations are rescaled accordingly, ensuring dimensional consistency and numerical stability of the coupled formulation.

\subsection{Thermodynamic lookup tables}

The lookup tables are calculated from Gibbs energy minimizations, i.e., thermodynamic equilibrium conditions. Thermodynamic calculations were carried out using ThermoLab \citep{vrijmoed2022} together with the thermodynamic dataset of \citet{holland2011}, with aqueous-fluid properties computed using the EOS of \citet{pitzer1994}. The potential solid phases considered were antigorite, brucite, talc, orthopyroxene, olivine, and quartz. The composition of all solid phases except quartz can vary with respect to the Fe/Mg ratio because \ce{Fe^2+} and \ce{Mg^2+} can occupy equivalent crystallographic sites. To account for these mixing effects, we used the solution models of \citet{evans2021} for all Mg-Fe silicates except brucite, for which we used an ideal mixing law. In the aqueous phase, dissolved silica is represented by the thermodynamic component $\mathrm{SiO_2}$(aq), i.e., a neutral monomeric dissolved silica component that captures the bulk dissolved silica content without resolving individual hydrated complexes.

It is important to note that the solubility of $\mathrm{FeO}$ and $\mathrm{MgO}$
in the fluid phase, which is based on interaction with the solid phases, is neglected for simplicity. Consequently, the thermodynamic–equilibrium
calculations yield $C^{f,\mathrm{MgO}} = 0$ and $C^{f,\mathrm{FeO}} = 0$. These quantities therefore do not enter the closure relations and are omitted from the lookup tables represented by the multivariate functions \eqref{eq:1rpw_LUT1}-\eqref{eq:1rpw_LUT4}.


In this study, phase–equilibrium calculations are performed with ThermoLab for a fluid-saturated system in which the lookup tables are parameterized by solid-only compositions, $\Tilde{C}^{s,SiO_2}$ and $\Tilde{C}^{s,MgO}$, expressed as normalized mass concentrations of the independent components $\mathrm{SiO_2}$ and $\mathrm{MgO}$:
\begin{equation}
    \Tilde{C}^{s,SiO_2} = \frac{C^{s,SiO_2}}{C^{s,SiO_2}+ C^{s,MgO} + C^{s,FeO}}
    \label{eq:dry_Cs_tilda_SO2}
\end{equation}
and 
\begin{equation}
    \Tilde{C}^{s,MgO} = \frac{C^{s,MgO}}{C^{s,SiO_2}+ C^{s,MgO} + C^{s,FeO}}
    \label{eq:dry_Cs_tilda_MgO}
\end{equation}

Our transport solver evolves variables for the full (wet) system and therefore tracks (i) the total mass of $\mathrm{SiO_2}$ per unit bulk volume, $\rho\,C^{t,\mathrm{SiO_2}}$, which includes both solid-hosted and dissolved silica, and (ii) the mass of dissolved silica in the fluid phase per unit bulk volume, $\rho^f\varphi\,C^{f,\mathrm{SiO_2}}$. To obtain the solid composition required by the lookup tables, we first compute the mass of $\mathrm{SiO_2}$ residing in the solid per unit bulk volume as
$\rho\,C^{t,\mathrm{SiO_2}}-\rho^f\varphi\,C^{f,\mathrm{SiO_2}}$.
We then normalize it by the total solid mass represented in ThermoLab-compatible description, which is the sum of the mass of $\mathrm{SiO_2}$ residing in the solid per unit bulk volume and the immobile (non-reacting) part of the solid matrix. In the considered system, the non-reacting solid components are $\mathrm{MgO}$ and $\mathrm{FeO}$, and we define their combined mass concentration as $C^{s,im}=C^{s,\mathrm{MgO}}+C^{s,\mathrm{FeO}}$, corresponding to an immobile-solid mass per unit bulk volume of $\rho^s(1-\varphi)\,C^{s,im}$. Hence, the dry-system solid mass concentration of $\mathrm{SiO_2}$ used in the solver takes the form:
\begin{equation}
    \Tilde{C}^{s,SiO_2} = \frac{\rho C^{t,SiO_2} - \rho^f \varphi C^{f,SiO2}}{\rho C^{t,SiO_2} - \rho^f \varphi C^{f,SiO2} + \rho^s (1-\varphi) C^{s,im}},
    \label{eq:dry_Cs_SO2} 
\end{equation}

Similarly, the dry-system solid mass concentration of $\mathrm{MgO}$ follows by normalizing the solid $\mathrm{MgO}$ mass, $\rho^s(1-\varphi)\,C^{s,\mathrm{MgO}}$, by the same dry-system total mass, yielding
\begin{equation}
    \Tilde{C}^{s,MgO} = \frac{\rho^s (1-\varphi) C^{s,MgO}}{\rho C^{t,SiO_2} - \rho^f \varphi C^{f,SiO2} + \rho^s (1-\varphi) C^{s,im}}.
    \label{eq:dry_Cs_MgO}
\end{equation}
This ensures full compatibility between the transport and thermodynamic solvers.


\subsection{Transport solver}

To solve the coupled system of governing equations, we employ an explicit finite difference scheme combined with an iterative, matrix-free pseudo-transient method \citep{frankel1950convergence, rass2022} on a rectangular space-time grid. Spatial discretization is performed using a conservative staggered-grid formulation \citep{virieux1986p}, which is equivalent to a finite-volume approach \citep{dormy1995numerical} and ensures local conservation of mass and momentum. This approach distinguishes between pseudo-time iterations, employed for the pressure and velocity solvers, and physical time stepping (e.g., phase densities, mass fractions per unit volume, and related physical quantities), thereby ensuring numerical stability and efficient convergence of the strongly coupled HMC system. Thermodynamic closure quantities required at each iteration are obtained by multidimensional interpolation of the precomputed lookup tables using the \texttt{interpn()} function in \textsc{MATLAB}.



\subsection{Modeling setup}

\begin{figure}
    \centering
    \includegraphics[width=12cm]{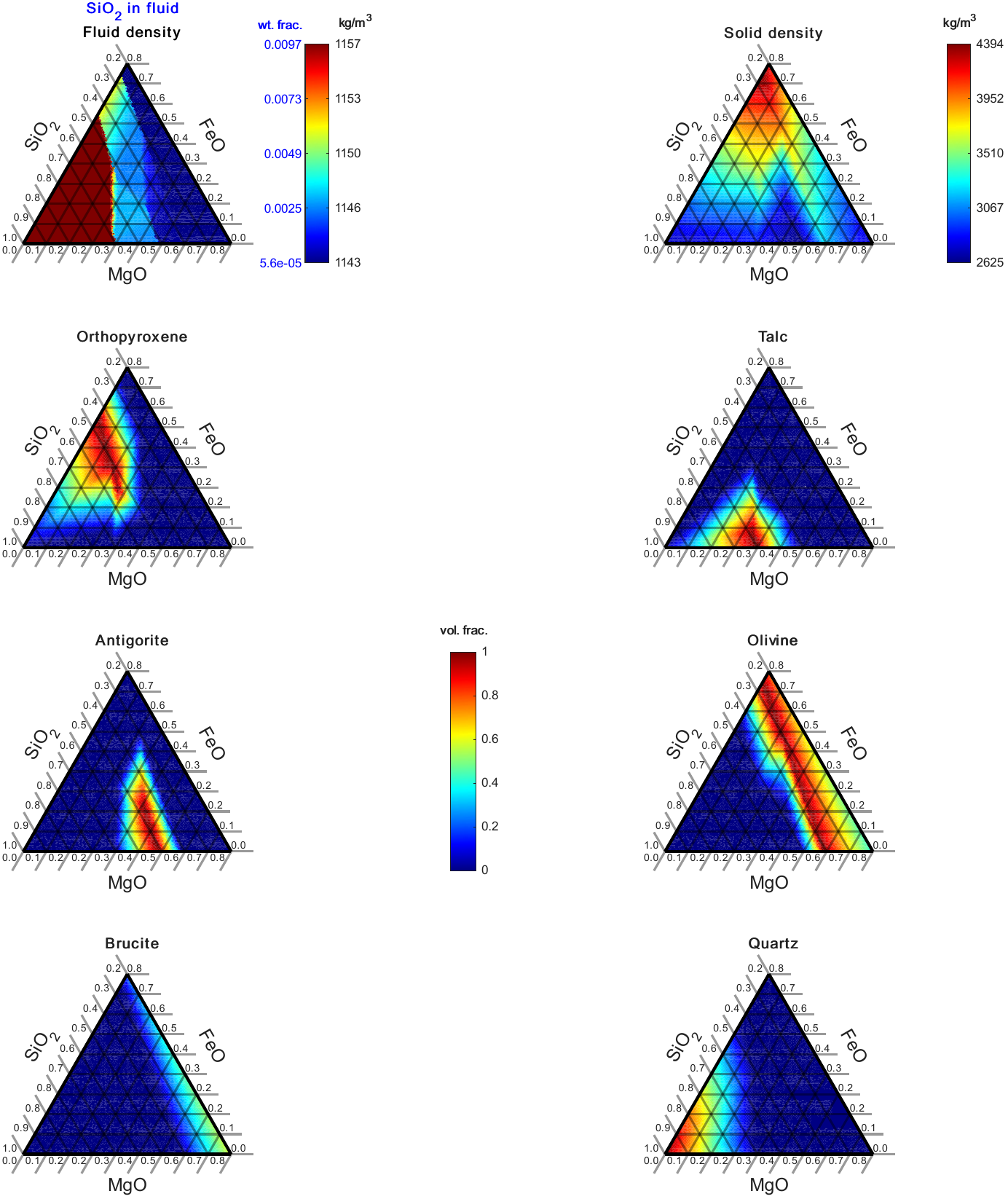}
    \caption[]{Representation of thermodynamic lookup tables evaluated at $\SI{550}{\degreeCelsius}$ and $\SI{3}{\giga\pascal}$ using Thermolab \citep{vrijmoed2022}. Fluid and solid phase densities, fluid composition, and the volume fractions of orthopyroxene, talc, antigorite, olivine, brucite, and quartz are shown in ternary diagrams as functions of the solid-phase concentrations of $\mathrm{SiO_2}$, $\mathrm{MgO}$, and $\mathrm{FeO}$ in the dry system. Fluid density and fluid composition exhibit identical topological structure in composition space; corresponding values are indicated by the color bars. The ternary diagrams of mineral volume fractions share a common color bar to facilitate direct comparison between phase abundances.}
    \label{fig:LUT}
\end{figure}

\subsubsection{Thermodynamic data}

Figure~\ref{fig:LUT} illustrates representative values extracted from the thermodynamic lookup tables, including the composition of the aqueous fluid, the bulk solid, and the modal abundances of orthopyroxene, talc, antigorite, and olivine, evaluated at constant $\SI{550}{\degreeCelsius}$ and $\SI{3}{\giga\pascal}$. 
The bulk composition at each lookup-table node is represented in ternary diagrams, illustrating the distribution of phase assemblages over the relevant compositional space.

The lookup tables were computed on a five-dimensional grid defined by pressure, temperature, and bulk composition of the system. Pressures range from 0.5 to \GPa{4.0} and temperatures from 300 to \cel{700} in the full \ce{FeO}-\ce{MgO}-\ce{SiO2} (FMS) composition space. It was considered that excess \ce{H2O}   was present to ensure the formation of the maximum hydrated mineral composition in the rock. 

Figure  \ref{fig:LUT} shows that at constant pressure, the fluid density depends mainly on the dissolved silica concentration ($C^{f,\ce{SiO2}}$) and therefore follows the topology of the fluid composition. Both quantities vary primarily with the bulk $\mathrm{SiO_2}$ content and display two plateaus that correspond to fluid compositions constrained by brucite at low $\mathrm{SiO_2}$ and by quartz at high $\mathrm{SiO_2}$ contents of the system. Solid density correlates with the solid $\mathrm{H_2O}$ content because anhydrous phases are denser than hydrous ones.

\subsubsection{Model configuration}


\begin{figure}
    \centering
    \includegraphics[width=18cm]{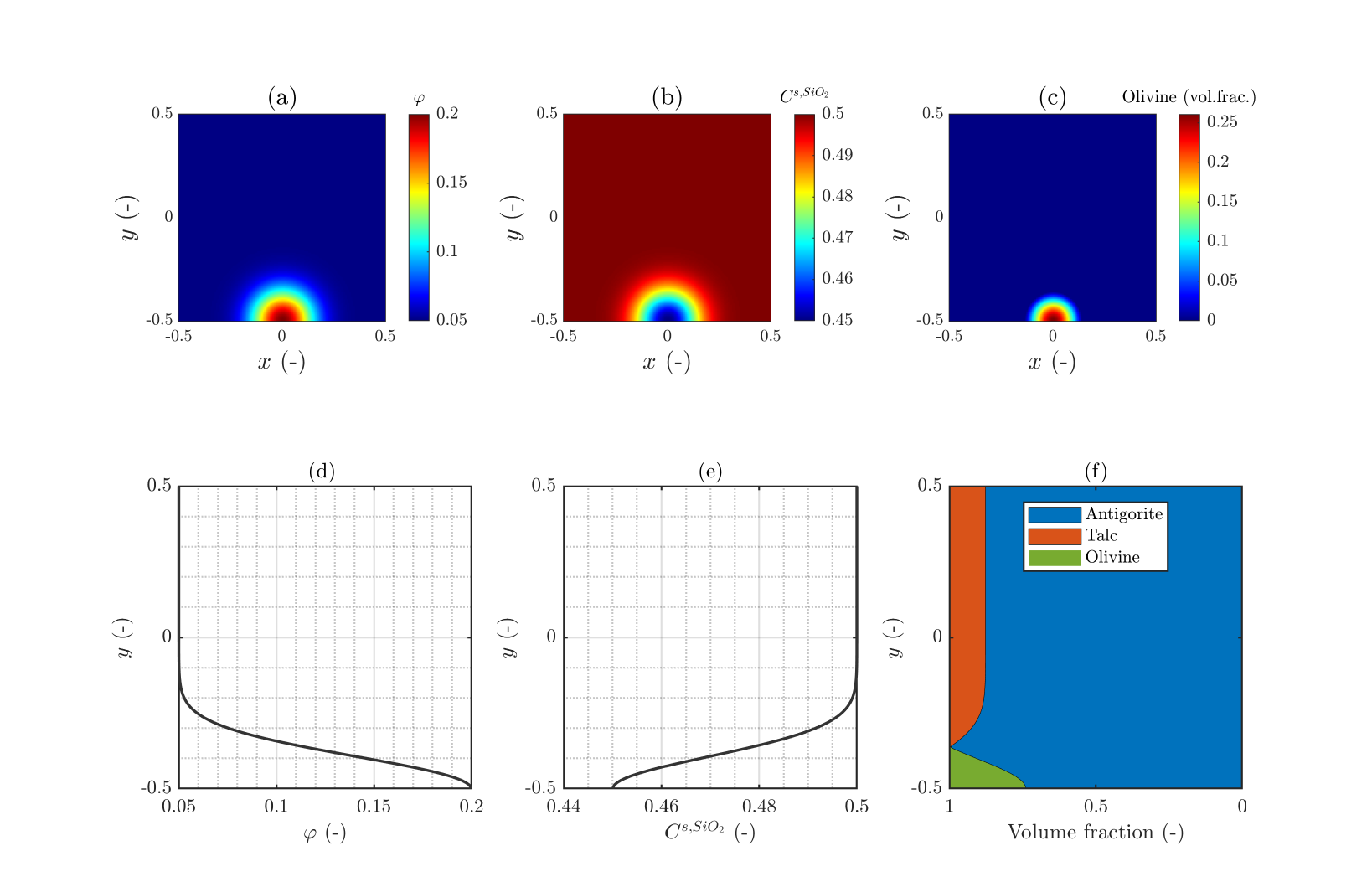}
    \caption[]{HMC model configuration (initial condition): Gaussian-shaped perturbations in porosity and in the solid-phase silica concentration $C^{s,\mathrm{SiO_2}}$ are imposed at the lower boundary, representing the localized injection of an $\mathrm{H_2O}$–$\mathrm{SiO_2}$ fluid into the porous medium. The injected fluid is prescribed at a temperature of $\SI{550}{\degreeCelsius}$ and ascends through the domain under the influence of buoyancy.
    (a) Initial porosity distribution. (b) Initial distribution of $\mathrm{SiO_2}$ concentration in the solid matrix. (c) Initial distribution of olivine volume fraction. (d) Initial porosity distribution shown along the central vertical slice in the $y$-direction of the 2D model. (e) Initial distribution of $\mathrm{SiO_2}$ concentration in the solid matrix along a central vertical slice in the $y$-direction in the 2D model. (f) Initial mineral composition of the rock along the central vertical slice in the $y$-direction in the 2D model. The volume fractions of the corresponding mineral phases are shown along the central vertical slice in the $y$-direction of the computational domain.}
    \label{fig:Initial_cond}
\end{figure}
\begin{figure}
    \centering
    \includegraphics[width=18cm]{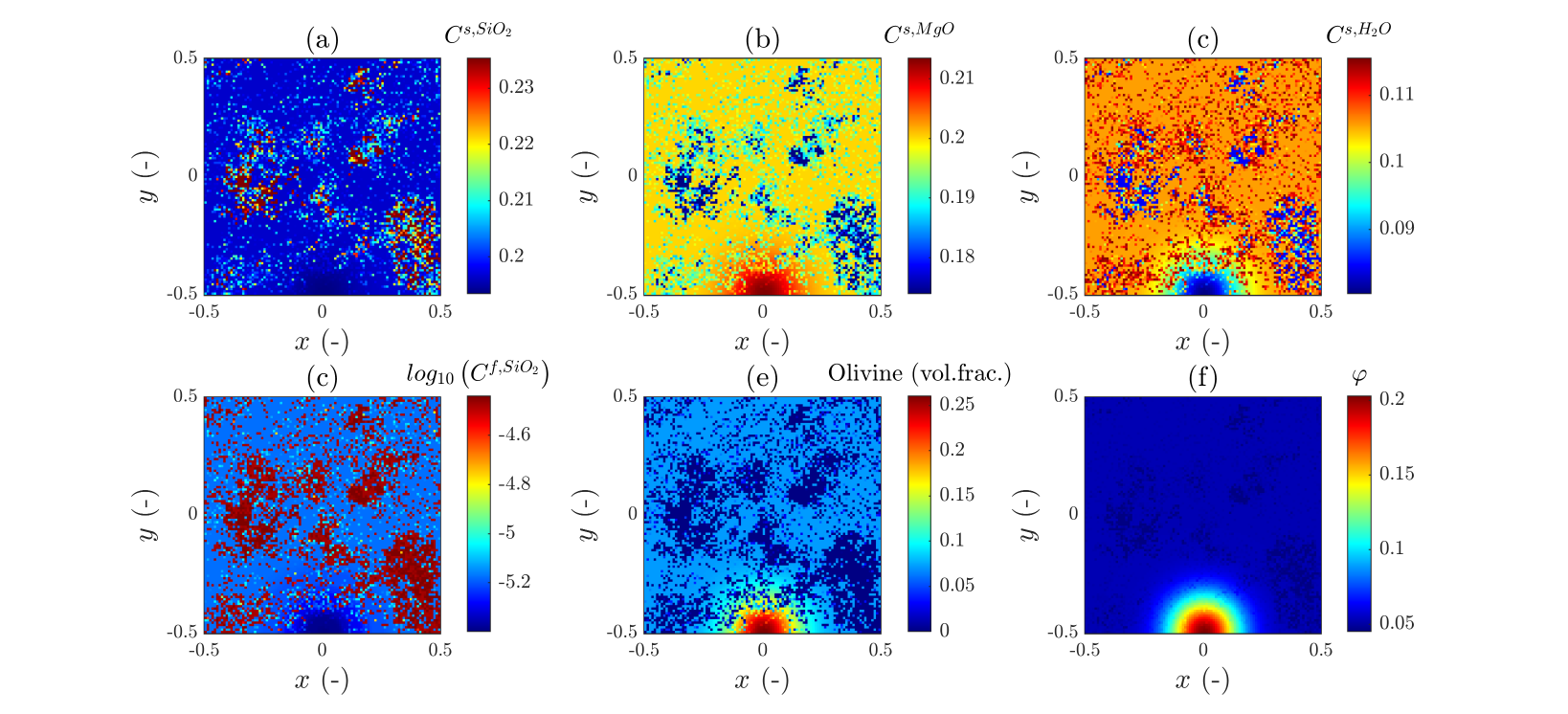}
    \caption[]{HMC model configuration using natural dataset based on compositional maps (cm-scale) of serpentinites from the Mirdita ophiolite (Albania) \citep{huber2024}: Gaussian-shaped perturbations in porosity and in the solid-phase silica concentration $C^{s,\mathrm{SiO_2}}$ are imposed at the lower boundary, representing the localized injection of an $\mathrm{H_2O}$–$\mathrm{SiO_2}$ fluid into the porous medium. The injected fluid is prescribed at a temperature of $\SI{550}{\degreeCelsius}$ and ascends through the domain under the influence of buoyancy. Initial distribution of 
    (a) $\mathrm{SiO_2}$ concentration in the solid matrix; (b) $\mathrm{MgO}$ concentration in the solid matrix; (c) fluid stored in the solid; (d) $\mathrm{SiO_2}$ concentration in the fluid; (e) volume fraction of olivine; (f) porosity.}
    \label{fig:Initial_cond_het}
\end{figure}

The model is formulated on a two-dimensional computational domain representing a fluid-saturated porous rock of prescribed chemical composition in thermodynamic equilibrium with the pore fluid ($\mathrm{H_2O}$--$\mathrm{SiO_2}$) at a temperature of $\SI{550}{\degreeCelsius}$ and a total pressure of $\SI{3.0}{\giga\pascal}$. In the following, we consider two complementary sets of simulations: (i) synthetic configurations with constant initial compositions of saturated rock that isolate the governing hydro-mechanical and chemical feedbacks, and (ii) natural configurations based on compositional maps of serpentinites of Mirdita ophiolite (Albania) that have not entered a subduction zone \citep{huber2024}, allowing direct geological interpretation.
The computational domain is a square box with a dimensionless extent of $101 \times 101$ in the horizontal ($x$) and vertical ($y$) directions, respectively, where $y$ denotes depth.

The initial porosity field is prescribed as Gaussian-shaped perturbations at the lower boundary of the domain, representing a localized fluid-rich source region or reservoir. This configuration introduces a spatially localized porosity anomaly that promotes focused fluid transport (Fig.~\ref{fig:Initial_cond}).

In addition to the porosity perturbation, the initial conditions include Gaussian-shaped perturbations in the solid-phase silica concentration $C^{s,\mathrm{SiO_2}}$, imposed at the lower boundary of the domain. Within this perturbed boundary region, the prescribed value of $C^{s,\mathrm{SiO_2}}$ is chosen such that it is in thermodynamic equilibrium with an incoming fluid of reduced silica content, but not in equilibrium with the surrounding saturated rock. This imposed chemical disequilibrium triggers mineral reactions and phase transformations at the boundary, thereby modeling the influx of a chemically undersaturated pore fluid and initiating reactive transport within the computational domain.

The system is subjected to a downward-pointing gravitational field. In the two-dimensional configuration, the domain is additionally affected by a background horizontal strike-slip shear deformation of magnitude comparable to buoyancy-driven forces, allowing for the investigation of coupled deformation, fluid flow, and reactions. The mechanical problem is solved with no-slip boundary conditions (zero velocity) on all domain boundaries. For the fluid-flow problem, the fluid pressure is set equal to the total pressure on all domain boundaries.


The numerical simulations are advanced for a total of $1000$ time steps, which is sufficient to capture the evolution of coupled hydro-mechanical-chemical processes. The dimensionless parameters employed in the simulations are selected to ensure numerical stability and efficient convergence of the pseudo-transient solution scheme.

In addition, we simulate the formation and propagation of reactive porosity waves for natural dataset. We use compositional maps of serpentinites from the Mirdita ophiolite (Albania) that has not entered a subduction zone \citep{huber2024}, which serve as initial conditions for the simulations. This approach allows us to reproduce the natural compositional heterogeneity and to investigate the HMC processes in natural datasets. The initial distributions of solid-phase $\mathrm{SiO_2}$, $\mathrm{MgO}$, $\mathrm{FeO}$, concentration, olivine volume fraction and porosity in the computational domain are shown in Fig.~\ref{fig:Initial_cond_het}.

\section{Results and discussion}

We present a systematic investigation of the coupled mechanical and chemical instabilities that arise during infiltration of a silica-undersaturated \ce{H_2O}-\ce{SiO2} fluid through a deforming and reacting porous rock with complex mineral composition.

\begin{figure}[H]
    \centering
    \includegraphics[width=16cm]{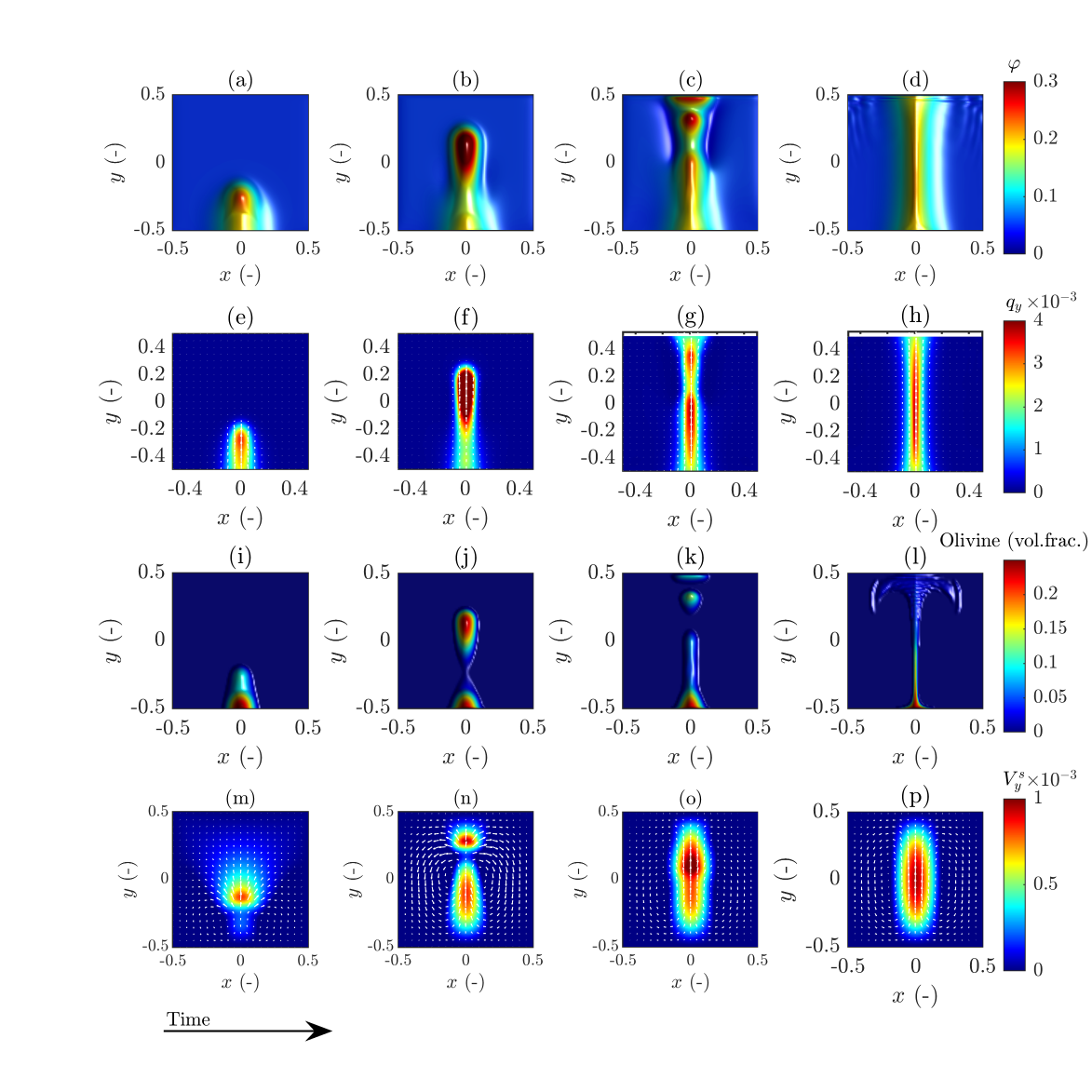}
    \caption[]{HMC modelling results at different time steps for the synthetic dataset, illustrating the evolution of reactive porosity waves driven by coupled (de)compaction and (de)hydration processes, resulting in the formation of characteristic channel-like structures in 2D. Thermodynamic data corresponds to serpentinites from the Mirdita ophiolite (Albania) that have not entered a subduction zone. The viscosity ratio is fixed to $\eta^{s}/\eta^{\varphi}=50$, the decompaction weakening $R=1000$. The initial model configuration includes Gaussian-shaped perturbations in porosity and in the solid-phase silica concentration $C^{s,\mathrm{SiO_2}}$ imposed at the lower boundary, representing localized injection of an $\mathrm{H_2O}$--$\mathrm{SiO_2}$ fluid into the porous medium and its subsequent upward migration under the influence of buoyancy. Each column corresponds to a different time step (1, 5, 15, 100), illustrating the temporal evolution of the simulated fields: (a--d) porosity $\varphi$; (e--h) vertical Darcy flux $q_y=\varphi\left(v_y^{f}-v_y^{s}\right)$; (i--l) olivine volume fraction; and (m--p) vertical solid velocity $v_y^{s}$.
}
    \label{fig:2D_eta_50}
\end{figure}

Figure~\ref{fig:2D_eta_50} summarizes the 2D HMC simulation results and demonstrates how reactive porosity waves develop and evolve under the coupled action of (de)compaction and dehydration reactions in the saturated mineral matrix. Mechanically, the imposed porosity perturbation triggers a hydro-mechanical instability that leads to compaction-driven focusing of fluid flow, as evidenced by the evolution of the vertical Darcy flux (Fig.~\ref{fig:2D_eta_50}e--h). At early times, this focusing manifests as a sequence of localized pulses; with ongoing evolution, additional pulses develop and interact, and the initially intermittent focusing progressively organizes into persistent channel-like, high-permeability transport pathways. This transition from pulsed focusing to channelized flow is reflected in the corresponding porosity evolution (Fig.~\ref{fig:2D_eta_50}(a-d). This behavior is consistent with earlier theoretical and numerical studies of porosity-wave dynamics and compaction-driven flow localization in deformable porous media \citep{connolly1998,connolly2007,yarushina2015a,rass2018,yarushina2022,schmalholz2026channelization}.

\begin{figure}[H]
    \centering
    \includegraphics[width=18cm]{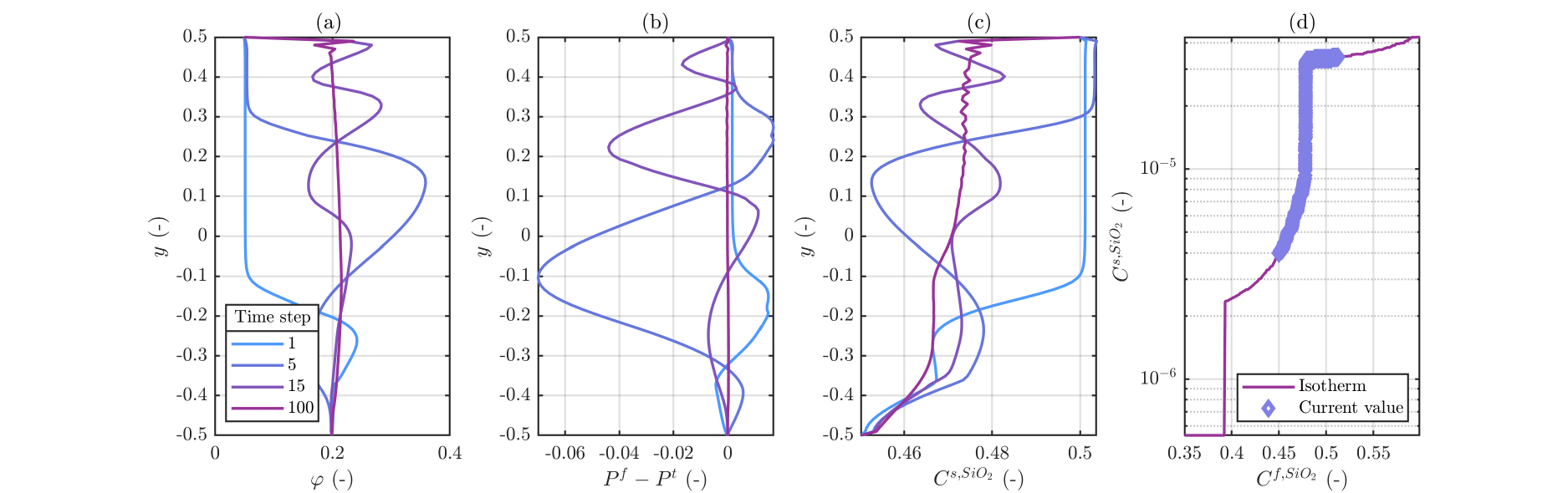}
    \caption[]{HMC modelling results for the synthetic dataset at different time steps (see legend) shown along the central vertical slice in the $y$-direction of the two-dimensional model (Fig.~\ref{fig:2D_eta_50}). (a) Porosity evolution $\varphi$; (b) pressure difference $P^{f}-P^{t}$; (c) solid-phase silica concentration $C^{s,\mathrm{SiO_2}}$; (d) isotherm relating the thermodynamic lookup-table data to the states observing during the reactive transport simulations. In panel (d), the purple curve denotes the equilibrium isotherm, expressing the relationship between $\mathrm{SiO_2}$ concentrations in the fluid and solid phases. Diamond markers indicate the states sampled during the simulations, and the arrow highlights the equilibrium relation corresponding to the composition of the fluid in the computational domain during simulation.
 }
    \label{fig:2D_sl}
\end{figure}

\begin{figure}[H]
    \centering
    \includegraphics[width=17cm]{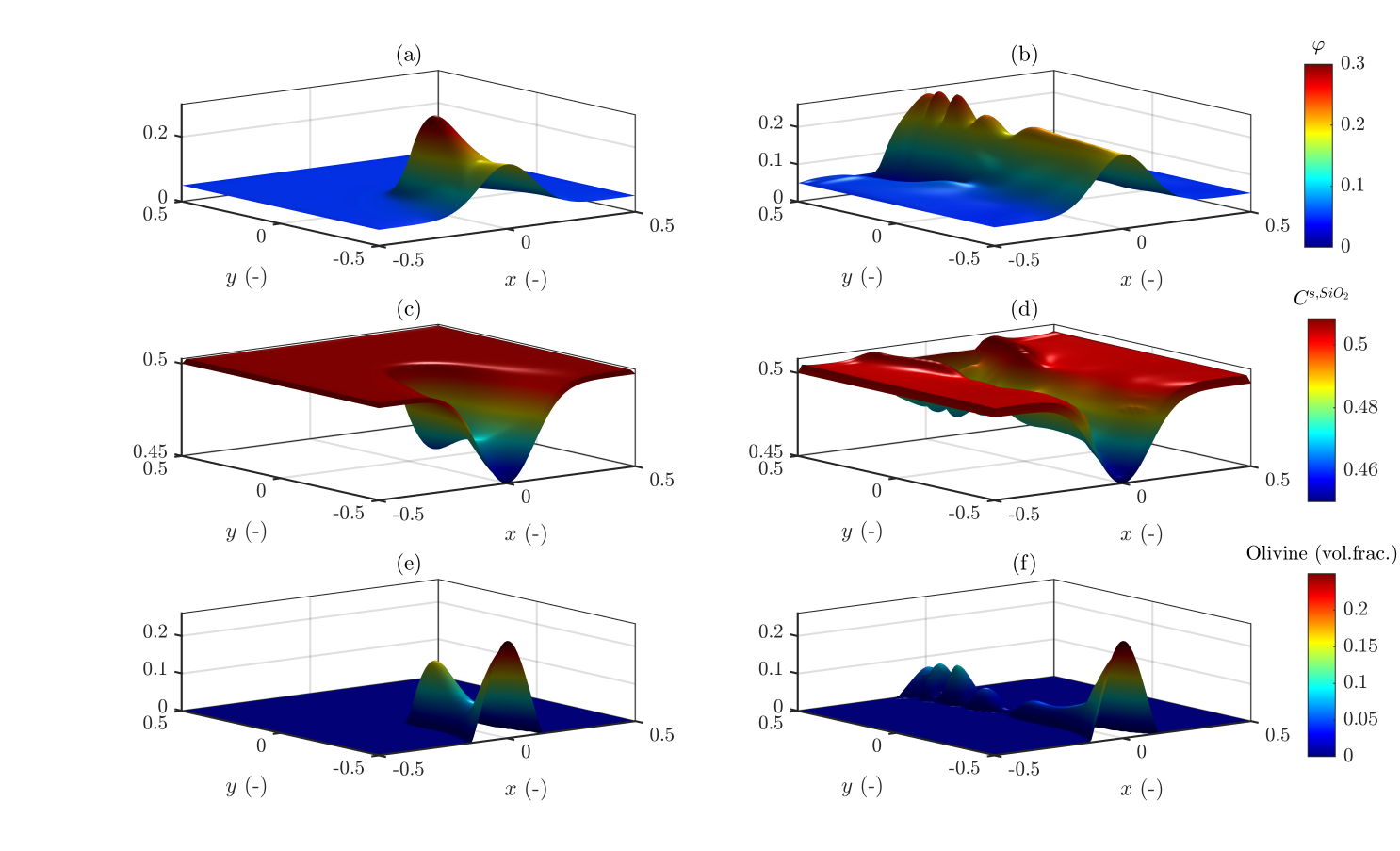}
    \caption[]{HMC modelling results for the synthetic dataset at different time steps in 2D for the serpentinites from the Mirdita ophiolite (Albania).  Each column corresponds to a different time step (2, 25), showing the temporal evolution of (a--b) porosity, (c--d) solid-phase silica concentration $C^{s,\mathrm{SiO_2}}$ and (e--f) olivine volume fraction. Infiltration of a fluid with reduced silica content decreases the solid-phase silica concentration $C^{s,\mathrm{SiO_2}}$ in the inflow region, thereby driving dehydration reactions and promoting olivine formation. The resulting olivine-rich zones correlate not only with regions of reduced solid silica content but also with zones of enhanced porosity and permeability, which emerge as mechanically induced (de)compaction channels that focus fluid transport.
    }
    \label{fig:2D_surf}
\end{figure}

Chemically, injection of a silica-undersaturated fluid drives dehydration reactions in the initially fluid-saturated mineral matrix, producing strong compositional contrasts and phase transformations. The resulting feedback between reaction-induced changes in solid composition (Fig. \ref{fig:2D_sl}(c)) and mechanically induced (de)compaction governs the initiation and evolution of reactive porosity waves (Fig. \ref{fig:2D_sl}(a)) and the associated localization patterns. Indeed, a major difference from previous hydro-mechanical models is that fluid flow in deforming rock is coupled to thermodynamically consistent mineral reactions using thermodynamic lookup tables verified with natural data \citep{huber2022}. Triggered dehydration reactions promote olivine formation (Fig. \ref{fig:2D_eta_50}(i-l)), and we demonstrate that olivine-rich zones spatially coincide with the high-porosity, high-permeability channels generated by mechanical (de)compaction (Fig. \ref{fig:2D_surf}). Predicted olivine formation patterns are consistent with previous reactive-transport studies \citep{plumper2017,huber2022, schmalholz2023}, but the present model additionally accounts for the coupling between mineral reactions, deformation, and flow localization, which our results show to be critical for controlling where olivine forms.

In all simulations, we consider the viscous limit of the Maxwell viscoelastic rheology for the solid skeleton, such that deformation is controlled by the shear viscosity $\eta^{s}$ and the bulk (compaction) viscosity $\eta^{\varphi}$ \citep{connolly1998,yarushina2023constitutive}.


\subsection{Systematic simulations}

The previous case shows that reactive porosity waves are inherently controlled by the sequence and strength of mechanical and chemical feedbacks. In the model, mechanical instabilities first induce focused fluid flow and the formation of high-permeability structures through (de)compaction. These mechanically generated flow pathways subsequently control the transport of a low--$\mathrm{SiO_2}$ fluid, which drives dehydration reactions and associated changes in solid composition. As a result, mechanical localization can either accelerate reactive porosity-wave development by sustaining fluid focusing and disequilibrium, or suppress it by redistributing stresses and closing reactive pathways. 

\begin{figure}[H]
    \centering
    \includegraphics[
        width=16cm,
        trim=0cm 0cm 0cm 0cm,
        clip
    ]{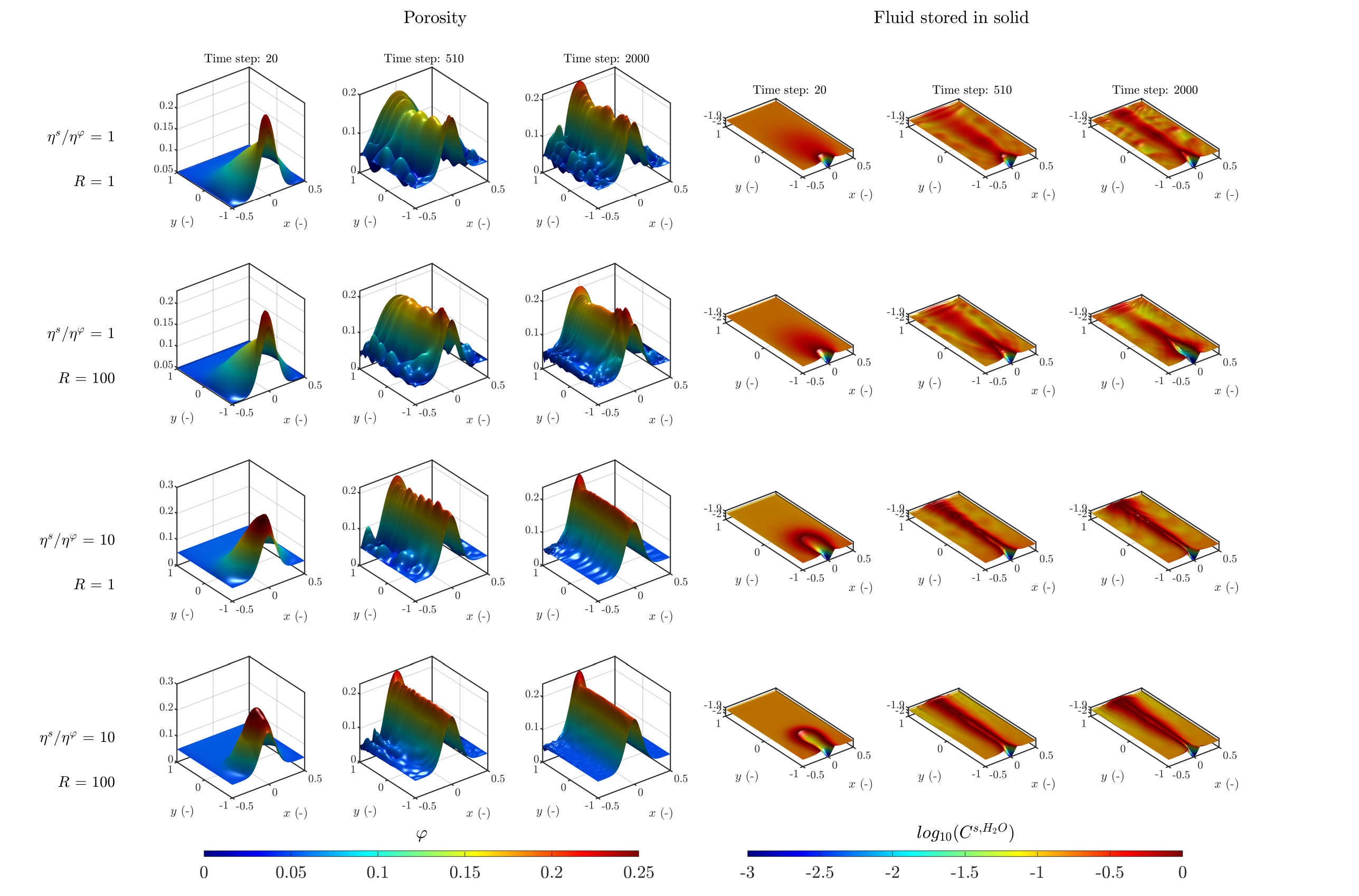}
    \caption{Results of systematic HMC simulations for the synthetic dataset for $\eta^{s}/\eta^{\varphi}\in\{1,10\}$ and $R\in\{1,100\}$, shown as snapshots at selected time steps (20, 510, 2000). Reactive porosity waves propagate in the $y$-direction, reflecting upward fluid migration and the associated localization of (de)compaction and reaction fronts. The panels are organized such that the viscosity ratio $\eta^{s}/\eta^{\varphi}$ and the decompaction weakening factor $R$ vary vertically, while time increases horizontally from left to right. The left block of panels shows the temporal evolution of porosity $\varphi$, whereas the right block of panels displays the corresponding distribution of water concentration in the solid phase, $log_{10}\left( C^{s,\mathrm{H_2O}} \right)$.}
\label{fig:2D_var1}

\end{figure}

\begin{figure}[H]
    \centering
    \includegraphics[
        width=14.5cm,
        trim=0cm 0cm 0cm 0cm,
        clip
    ]{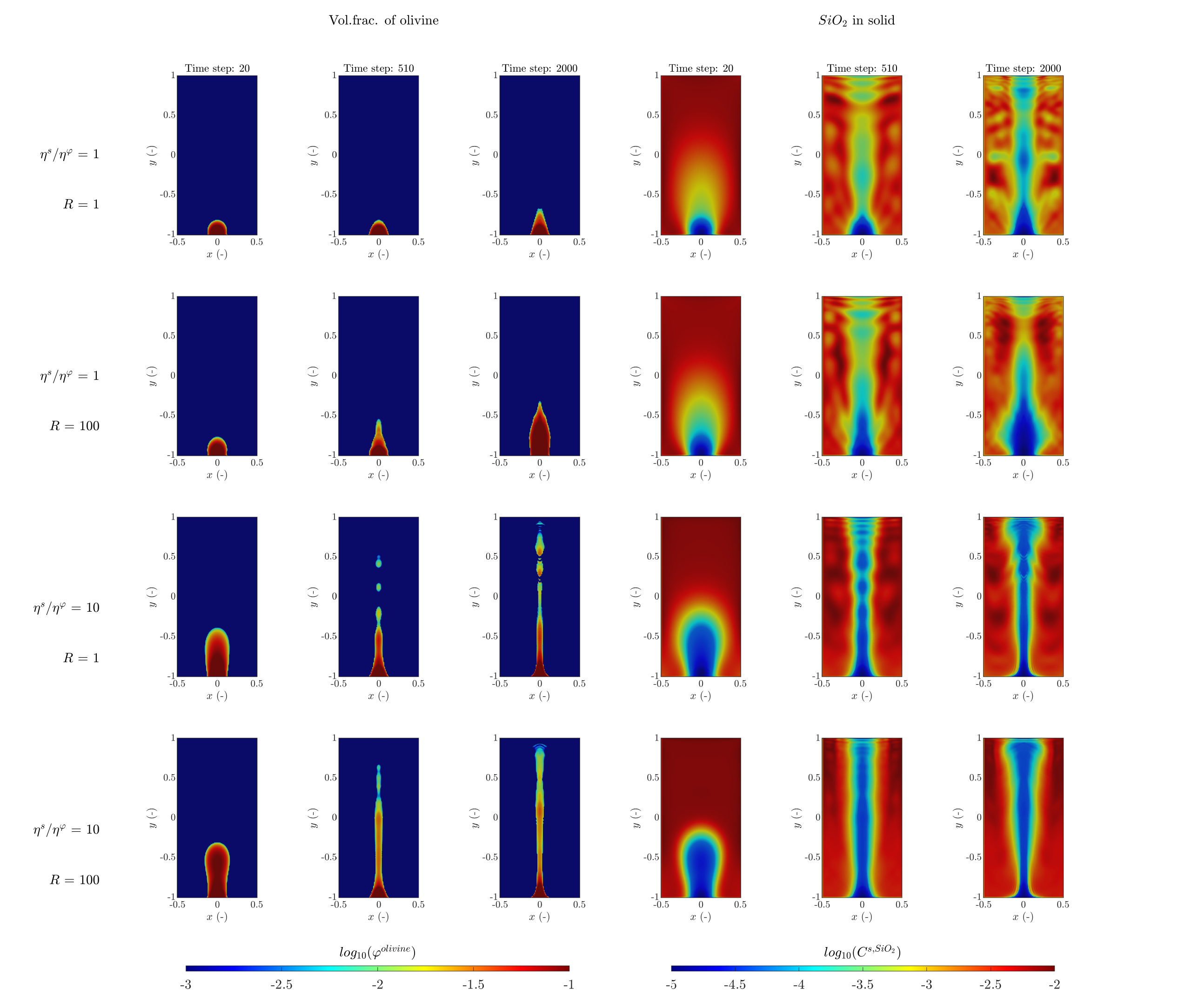}
    \caption{Results of systematic HMC simulations for the synthetic dataset for $\eta^{s}/\eta^{\varphi}\in\{1,10\}$ and $R\in\{1,100\}$, shown as snapshots at selected time steps (20, 510, 2000). Reactive porosity waves propagate in the $y$-direction. The panels are organized such that the viscosity ratio $\eta^{s}/\eta^{\varphi}$ and the decompaction weakening factor $R$ vary vertically, while time increases horizontally from left to right. The left column shows the temporal evolution of volume fraction of olivine, whereas the right column displays the corresponding distribution of $\mathrm{SiO_2}$ concentration in the solid phase, $log_{10}\left( C^{s,\mathrm{SiO_2}} \right)$.}
    \label{fig:2D_var2}
\end{figure}

To investigate it, we perform systematic parameter variations to determine how viscosity ratio $\eta^{s}/\eta^{\varphi}$ and the decompaction weakening factor $R$ jointly control the initiation, propagation, and morphology of reactive porosity waves in 2D HMC simulations. The results are summarized in Figs.~\ref{fig:2D_var1} and \ref{fig:2D_var2}, which document the time evolution of porosity, water content in the solid, solid-phase $\mathrm{SiO_2}$ concentration, and olivine volume fraction for $\eta^{s}/\eta^{\varphi}\in\{1,10\}$ and $R\in\{1,100\}$.

The viscosity ratio $\eta^{s}/\eta^{\varphi}$ quantifies the relative resistance of the solid skeleton to shear deformation versus volumetric (de)compaction, while the decompaction weakening factor $R$ accounts for weakening during dilation by reducing the effective bulk viscosity in decompaction \citep{connolly1998, connolly2007, rass2018, yarushina2020model}. Similar to earlier analyses of decompaction weakening \citep{connolly2007,rass2019}, increasing $\eta^{s}/\eta^{\varphi}$ and $R$ promotes stronger localization of porosity anomalies and more pronounced fluid focusing (Fig.~\ref{fig:2D_var1}), consistent with the systematic study of \citet{wang2022physics}. This enhanced focusing is reflected in the distribution of water stored in the solid matrix, which evolves from a diffuse pattern into a channel-like structure aligned with the high-permeability pathways. At low $\eta^{s}/\eta^{\varphi}$ and weak decompaction weakening, fluid flow remains relatively diffuse.


In the coupled HMC framework, however, this focusing is not purely mechanical: localized fluxes sustain chemical disequilibrium, accelerate reaction progress, and feedback on porosity through reaction-induced changes in solid composition (Fig. \ref{fig:2D_var2}(b)) and phase assemblage (Fig. \ref{fig:2D_var2}(a)). Figure~\ref{fig:2D_var2} demonstrates that the mechanically generated high-permeability pathways organize the spatial distribution of reactive fronts. This effect is expressed by the localization of olivine formation, which systematically coincides with regions of reduced solid-phase $\mathrm{SiO_2}$ concentration. At low viscosity ratios $\eta^{s}/\eta^{\varphi}$ and weak decompaction weakening ($R$ close to unity), mechanical focusing remains limited and the reaction front is effectively blocked: the matrix undergoes only minor decomposition and olivine occurs predominantly as a diffuse, front-like pattern controlled by slow compositional redistribution rather than by focused advection. 

In contrast, at higher viscosity ratios and stronger decompaction weakening, reactive porosity waves acquire a channel-like topology. In this regime, pronounced reaction fronts develop within actively decompacting regions, and olivine forms preferentially along the same channelized structures that accommodate focused fluid flow.

\subsection{Natural dataset}

\begin{figure}[H]
    \centering
    \includegraphics[width=16cm]{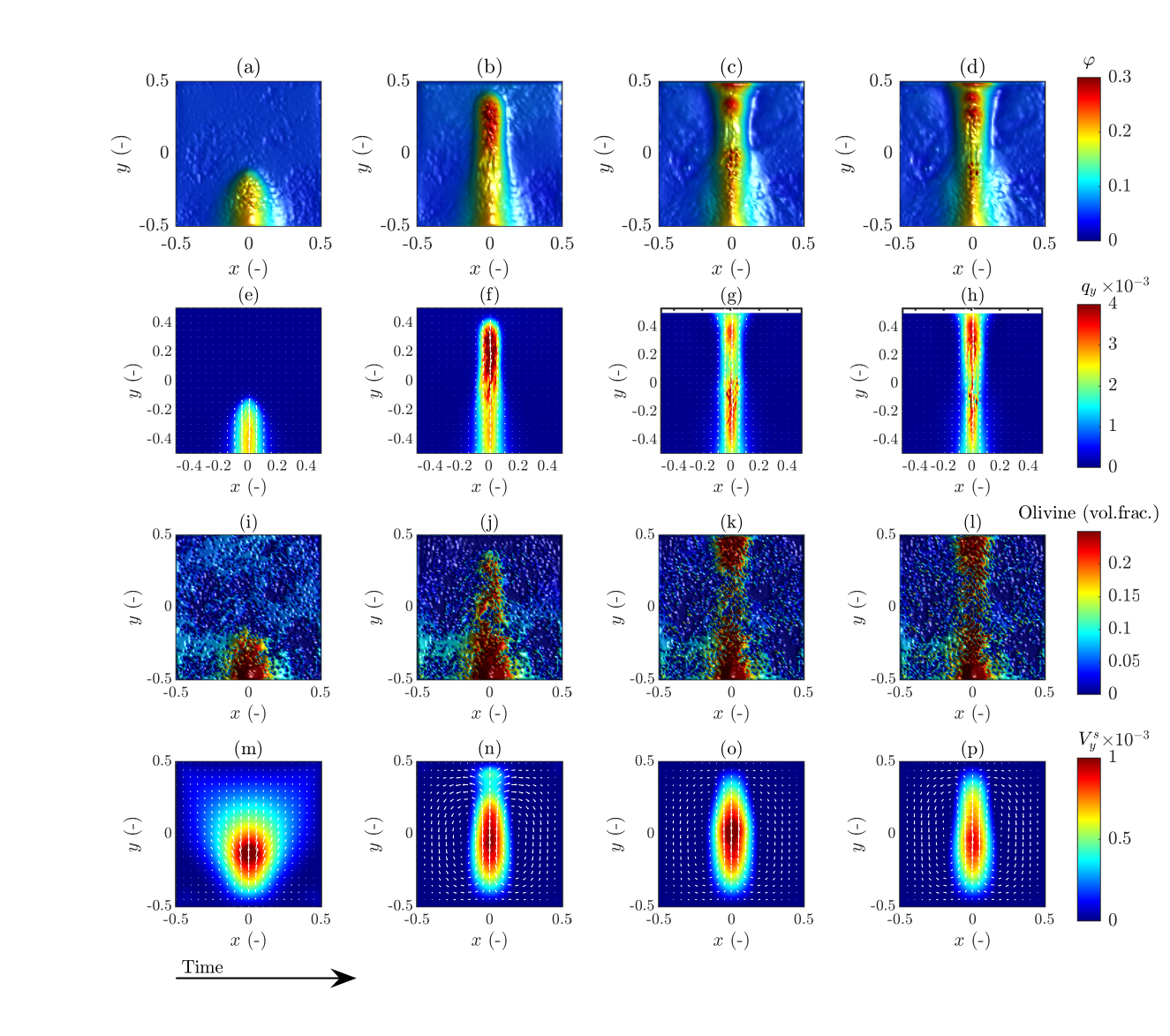}
    \caption[]{HMC modelling results at different time steps for the natural dataset: laboratory-measured cm-scale chemical maps of serpentinites from the Mirdita ophiolite (Albania) \citep{huber2024}. The viscosity ratio is fixed to $\eta^{s}/\eta^{\varphi}=50$. The initial model configuration includes Gaussian-shaped perturbations in porosity and in the solid-phase silica concentration $C^{s,\mathrm{SiO_2}}$ imposed at the lower boundary, representing localized injection of an $\mathrm{H_2O}$--$\mathrm{SiO_2}$ fluid into the porous medium and its subsequent upward migration under the influence of buoyancy. Each column corresponds to a different time step (1, 5, 15, 100), illustrating the temporal evolution of the simulated fields: (a--d) porosity $\varphi$; (e--h) vertical Darcy flux $q_y=\varphi\left(v_y^{f}-v_y^{s}\right)$; (i--l) olivine volume fraction; and (m--p) vertical solid velocity.
}
    \label{fig:2D_eta_50_het}
\end{figure}

\begin{figure}[H]
    \centering
    \includegraphics[width=17cm]{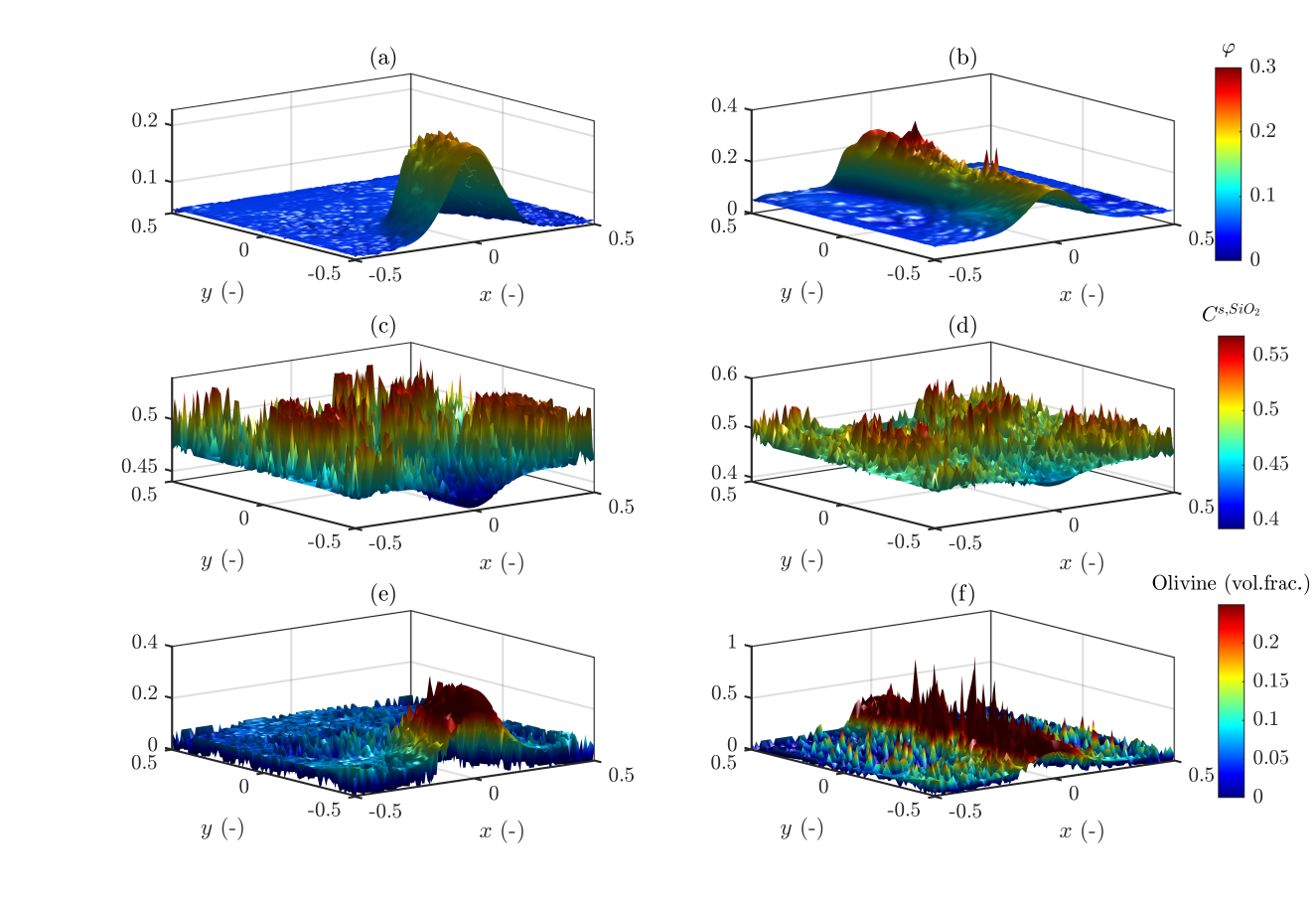}
    \caption[]{HMC modelling results at different time steps for the natural dataset: laboratory-measured cm-scale chemical maps of serpentinites from the Mirdita ophiolite (Albania) \citep{huber2024}. The viscosity ratio is fixed to $\eta^{s}/\eta^{\varphi}=50$. Each column corresponds to a different time step (2, 20), showing the temporal evolution of (a--b) porosity, (c--d) solid-phase silica concentration $C^{s,\mathrm{SiO_2}}$ and (e--f) olivine volume fraction. Infiltration of a fluid with reduced silica content decreases the solid-phase silica concentration $C^{s,\mathrm{SiO_2}}$ in the inflow region, thereby driving dehydration reactions and promoting olivine formation. The resulting olivine-rich zones correlate not only with regions of reduced solid silica content but also with zones of enhanced porosity and permeability, which emerge as mechanically induced (de)compaction channels that focus fluid transport.
    }
    \label{fig:2D_surf_het}
\end{figure}

In addition, we extend the synthetic model configurations to geologically relevant data describing the mineral composition of the slabs. Figures~\ref{fig:2D_eta_50_het}-\ref{fig:2D_surf_het} show the results of the 2D HMC simulations for the natural dataset, in which the initial model configuration is prescribed by the chemical maps of serpentinites from the Mirdita ophiolite \citep{huber2024}, shown in Fig.~\ref{fig:Initial_cond_het}. The overall dynamical behavior is qualitatively similar to the synthetic reference case (Fig. \ref{fig:2D_eta_50}): we observe the evolution of reactive porosity waves driven by coupled (de)compaction and (de)hydration processes, resulting in the formation of characteristic channel-like structures in 2D. These structures characterizes higher porosity and permeability and focus fluid flow. This behavior is consistent with permeability estimates for the Erro-Tobbio dehydration vein networks, which act as efficient drainage systems and allow rapid fluid percolation even at low bulk porosity \citep{arias2026quantifying}. In the considered cases, the reactive porosity waves are initiated in the silica-poorer region near the lower boundary and migrate through progressively silica-richer parts of the domain. The reactive porosity waves associated with the focused fluid flow with lower silica content interact with spatial variations in mineral assemblage and bulk composition, which leads to a strongly nonuniform reaction progress and phase redistribution. As it traverses the domain, the waves drive a sequence of dehydration reactions and associated phase transformations \citep{huber2024}. In particular, along compositional gradients, the wave promotes the progressive conversion of talc to antigorite and, ultimately, the formation of olivine (Fig. \ref{fig:2D_surf_het}(e,f)).

This behavior establishes a direct link between the numerical model and geological processes of fluid-induced metasomatism and reaction-driven channelization. The incoming fluid is initially silica-undersaturated with respect to the surrounding rock, and reactive transport proceeds through dehydration reactions and silica dissolution within localized channel-like structures that develop in the fluid-saturated rock as a result of mechanical decompaction.

The evolution of solid silica content further supports this interpretation. Fig.~\ref{fig:2D_surf_het}(c,d) shows that the solid $\mathrm{SiO_2}$ concentration decreases as the porosity waves migrate through the domain, representing progressive silica dissolution and its migration from the solid matrix to the fluid along the focused flow pathways. Within the silica-richer region, orthopyroxene is the dominant stable phase, accompanied by minor talc. Interaction with the low-silica fluid drives olivine formation primarily via the reaction of orthopyroxene to olivine and, to a lesser extent, through the reaction sequence talc to antigorite to olivine. 

\begin{figure}[H]
    \centering
    \includegraphics[width=16cm]{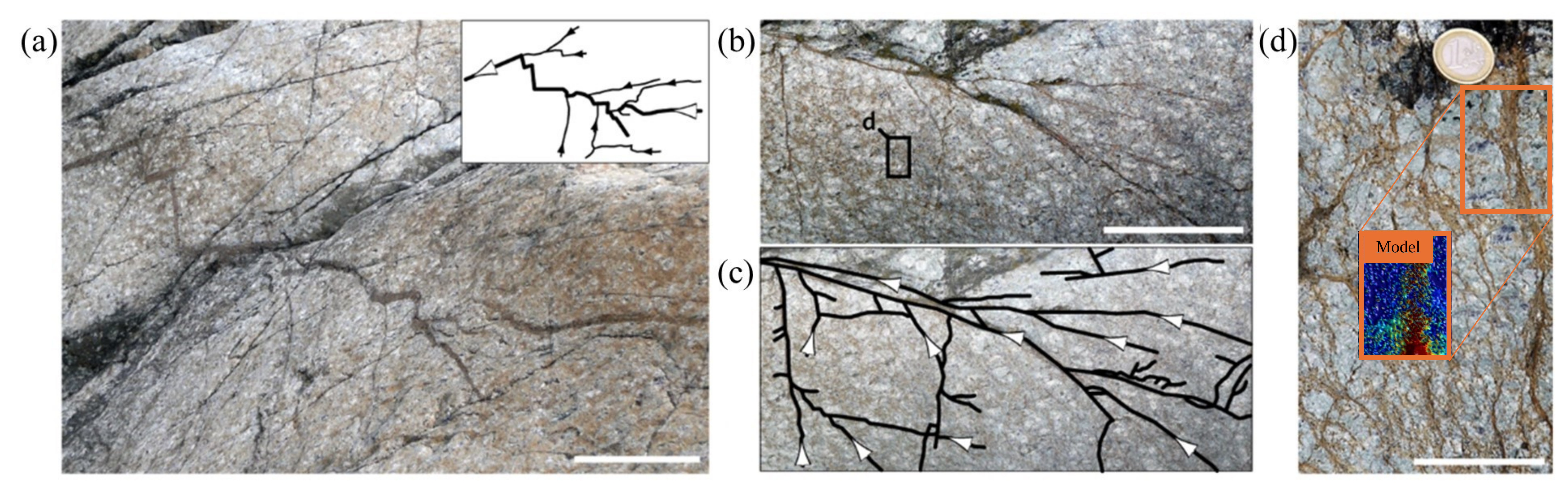}
    \caption[]{Photographs of eclogite-facies olivine vein networks produced by serpentinite dehydration in the Erro-Tobbio area, Italy (modified from \citep{plumper2017}). Vein formation during dehydration is observed ubiquitously across scales, as illustrated by successive zoom-ins from left to right. The inset in (a) and panel (c) highlight the channelized nature of the vein network; arrows indicate inferred preferential fluid-escape pathways. Scale bars: (a,b) $50\,\mathrm{cm}$; (d) $5\,\mathrm{cm}$; panels (b) and (c) are shown at the same scale. Panel (d) highlights the correspondence between the modeled dominant olivine-rich channel and the dominant channel observed in natural samples.
}
    \label{fig:vein_example}
\end{figure}

Notably, our model predicts the formation of a dominant olivine-rich, fluid-rich channel (Figs.~\ref{fig:2D_eta_50}, \ref{fig:2D_var2}, and \ref{fig:2D_eta_50_het}), representing a local segment (building block) of the broader vein-network architecture observed in nature. The spatial organization and first-order characteristics of this major channel are consistent with reaction-driven channelization documented in hydrated serpentinites from the Erro-Tobbio area, Italy (Fig.~\ref{fig:vein_example}; \citealp{scambelluri1997high,plumper2017,schmalholz2023,arias2026quantifying}).

\conclusions  
We propose a hydro-mechanical-chemical (HMC) 2D model coupled with thermodynamic data corresponding to serpentinites from the Mirdita ophiolite (Albania). By incorporating laboratory-based element distribution maps as initial conditions, the model captures realistic mineralogical heterogeneity and enables direct comparison with natural systems.
Our numerical results demonstrate the spontaneous formation of fluid-rich, high-permeability channel within a deforming and dehydrating serpentinite matrix during infiltration of a low--$\mathrm{SiO_2}$ $\mathrm{H_2O}$--$\mathrm{SiO_2}$ fluid. Channelization emerges from the coupled feedback between mechanical decompaction, focused fluid flow, and reaction-driven changes in mineral assemblage. In particular, silica depletion of the solid matrix promotes dehydration reactions and progressive olivine formation, which spatially correlates with zones of elevated porosity and permeability. These channel-like structure evolve from initially localized reactive porosity waves and organize into persistent transport pathways that efficiently focus fluid migration.
Preliminary numerical simulations reproduce the first-order characteristics of a dominant olivine-rich channel comparable to those observed at Erro Tobbio, suggesting that coupled hydro–mechanical–chemical feedbacks can generate such localized pathways during dehydration. While the present modeling results do not show the full branching vein-network architecture, the results document the dynamic formation of fluid-rich structures within an otherwise low-porosity rock matrix and support a unifying interpretation in which larger-scale channel networks may emerge from the repeated development and interaction of such dominant channel segments across scales. Such networks provide an efficient and dynamically sustained mechanism for fluid extraction and transport in hydrated lithologies and are therefore likely to play a key role in regulating fluid release, metasomatism, and mass transfer in subduction-zone environments. By explicitly accounting for dissolved $\mathrm{SiO_2}$ in the infiltrating fluid, our framework provides a consistent basis for linking fluid focusing in reacting and deforming host rocks during dehydration and metasomatism to the formation of quartz-rich veins, an ubiquitous indicator of silica transport in orogenic environments.

\clearpage

\begin{acknowledgements}
This research was funded by the Deutsche Forschungsgemeinschaft (DFG) through grant CRC 1114 Scaling Cascades in Complex Systems, Project Number 235221301, Project (C09) – "Dynamics of rock dehydration on multiple scales".
\end{acknowledgements}



\bibliographystyle{copernicus}
\bibliography{Reactive}






%
%

\clearpage

\end{document}